\documentclass[aps,
preprint,
superscriptaddress,
amsfonts,amsmath,amssymb]{revtex4-1}

\usepackage{graphicx}

\newcommand{\fref}[1]{Figure~\ref{#1}}

\usepackage{bbm}
\usepackage{color}

\begin{document}



\title{
Induced charges in a Thomas-Fermi metal: \\ insights from molecular simulations
}

\author {Swetha Nair}
\affiliation{Sorbonne Universit\'e, CNRS, Physicochimie des \'electrolytes et nanosyst\`emes interfaciaux, UMR 8234 PHENIX, F-75005, Paris, France}
\author{Giovanni Pireddu}
\affiliation{Sorbonne Universit\'e, CNRS, Physicochimie des \'electrolytes et nanosyst\`emes interfaciaux, UMR 8234 PHENIX, F-75005, Paris, France}
\author{Benjamin Rotenberg}
\email{benjamin.rotenberg@sorbonne-universite.fr}
\affiliation{Sorbonne Universit\'e, CNRS, Physicochimie des \'electrolytes et nanosyst\`emes interfaciaux, UMR 8234 PHENIX, F-75005, Paris, France}
\affiliation{R\'eseau sur le Stockage Electrochimique de l'Energie (RS2E), FR CNRS 3459, 80039 Amiens Cedex, France}

\begin{abstract}
We study the charge induced in a Thomas-Fermi metal by an ion in vacuum, using an atomistic description employed in constant-potential molecular dynamics simulations, and compare the results with the predictions from continuum electrostatics. Specifically, we investigate the effects of the Thomas-Fermi screening length $l_{TF}$ and the position $d$ of the ion with respect to the surface on the induced charge distribution in a graphite electrode. The continuum predictions capture most of the features observed with the atomistic description (except the oscillations due to the atomic sites of the graphite lattice), provided that $d$ and $l_{TF}$ are larger than the inter-atomic distances within the electrode. At large radial distance from the ion, the finite $l_{TF}$ case can be well approximated by the solution for a perfect metal using an effective distance $d+l_{TF}$. This requires a careful definition of the effective interface between the metal and vacuum for the continuum description. Our atomistic results support in particular an early analytical prediction [Vorotyntsev and Kornyshev, \emph{Zh. Eksp. Teor. Fiz.}, 1980, {\bf 78}(3), 1008] for a single charge at the interface between a Thomas-Fermi metal and a polarizable medium, which remains to be tested in atomistic simulations with an explicit solvent.
\end{abstract}

\maketitle

\section{Introduction}

The polarization of a metallic surface by an electrolyte solution, which plays a key role in electrochemistry or for the electrochemical storage of energy, results from the interplay of the charge distributions on both sides of the interface. Firstly, a polar solvent screens the electrostatic interactions between ions, which in a continuum picture is simply reduced by a factor equal to the solvent relative permittivity $\epsilon_r$. The distance over which this screened interaction between two elementary charges $e$ is equal to the thermal energy, is the so-called Bjerrum length $l_B=e^2/4\pi\epsilon_0\epsilon_rk_BT$, with $\epsilon_0$ the vacuum permittivity, $k_B$ Boltzmann's constant and $T$ the temperature ($l_B\approx 7$~\AA\ in water at room temperature). Secondly, the combined effect of thermal motion and electrostatic interactions between ions results in ionic screening. In dilute electrolytes, Debye-H\"uckel theory predicts that the characteristic length for the decay of electrostatic interactions is the Debye screening length, which depends on the Bjerrum length, as well as the concentrations and the valencies of the ions. Finally, in metals the screening of the electric field by electrons can be described \emph{e.g.} by Lindhardt theory, which reduces in the long-wavelength limit to the Thomas-Fermi model of screening~\cite{thomas1927a,fermi1927a}. The characteristic length for the decay inside the metal is the Thomas-Fermi screening length, $l_{TF}$, which depends on the density of states at the Fermi level~\cite{AshcroftMermin}.

Screening inside the metal has important consequences on the charge induced by an external charge distribution, on electrostatic interactions between ions near the surface and in turn on the structure of the electric double layer, on the capacitance of the interface and more generally on interfacial thermodynamics and kinetics. These aspects have been successfully investigated in the framework of continuum (in particular non-local) electrostatics~\cite{kornyshev_image_1977,kornyshev1978a,vorotyntsev_1980,kornyshev1982a,dzhavakhidze_activation_1987,phelps1990a,luque2012a,kornyshev2013a,rochester2013a,comtet2017a,Kaiser_2017}. In particular, Kornyshev and co-workers showed that the potential arising from a point charge at the interface between a dielectric medium (with permittivity $\epsilon_1$) and Thomas-Fermi metal (with dielectric function $\epsilon(k)=\epsilon_2\times(1+k_{TF}^2/k^2)$, where $k_{TF}=l_{TF}^{-1}$ is the reciprocal of the Thomas-Fermi screening length and $k$ the wavevector) decays laterally with a characteristic length $\propto \epsilon_1 l_{TF} / \epsilon_2$, \emph{i.e.} involving the properties of both materials. In parallel, the response of the electronic distribution of metals has been analyzed using electronic Density Functional Theory (DFT), initially explored in 1D geometries~\cite{Lang-jellium-1973,schmickler_interphase_1984,Smith-jellium-1989}, with subsequent studies incorporating more advanced functionals and atomically resolved surfaces~\cite{jung_self-consistent_2007,yu_abinitio_2008}, particularly within the framework of the Jellium (uniform electron gas) model. In order to link such descriptions with the continuum picture, one needs in particular to address the question of where to locate the interface, so-called ``Jellium edge'', between the metal and the external medium (vacuum or liquid).

From the simulation point of view, efficient algorithms have also been introduced to capture the mutual polarization of media with different permittivities in the presence of ions, to investigate the properties of interfacial electrolytes described by explicit ions in an implicit solvent~\cite{tyagi_icmmm2d_2007,tyagi_electrostatic_2008,arnold2013a,girotto_simulations_2017,nguyen_incorporating_2019,maxian_fast_2021,yuan_particleparticle_2021,dos_santos_modulation_2023}. \emph{Ab initio}~\cite{sakong_electric_2018,elliott_qmmd_2020,le_modeling_2021,sakong_structure_2022,takahashi_accelerated_2022,sundararaman_improving_2022,grisafi_predicting_2023} and classical~\cite{Siepmann_1995,reed2007a,Reed_2008,fedorov2014a,onofrio_voltage_2015,breitsprecher_electrode_2015,liang_applied_2018,nakano_chemical_2019,takahashi_unified_2022,sitlapersad_simple_2024,scalfi_molecular_2021,jeanmairet_microscopic_2022} molecular dynamics simulations have also become an essential tool to investigate electrode/electrolyte interfaces and electrochemical systems taking into account the atomic structure of the metal and the molecular features of the electrolyte. The charge induced inside the metal by the electrolyte plays a role on the interfacial structure, dynamics and thermodynamics~\cite{willard2009a,limmer2013b,
willard2013a,paek2015a,buraschi2020a,serva_effect_2021,loche_effects_2022,ntim_role_2020,ntim_molecular_2023,ntim_differential_2024}; the statistics of its fluctuations can further be used as a powerful probe of the latter~\cite{limmer2013a,merlet_electric_2014,scalfi_charge_2020,Pireddu_prl_2023}. Some strategies have also been introduced to describe Thomas-Fermi metals in such molecular simulations, which are able to capture screening within the metal and its consequences on interfacial thermodynamics~\cite{Scalfi_tfmodel_2020,schlaich_electronic_2021,scalfi-pnas-2021}.

Recently, we investigated the charge induced in a perfect metal ($l_{TF}=0$) by a single ion near a model gold electrode using classical molecular simulations in the constant-potential ensemble, focusing on the effects of the atomic lattice of the solid and of a molecular solvent (water), compared to an ion in vacuum and with analytical predictions from continuum electrostatics~\cite{pireddu_molecular_2021}. In the present work, we consider instead the effects of screening inside the solid (tuned via the Thomas-Fermi length $l_{TF}$) on the charge induced by an ion in vacuum. We address in particular the effects of periodic boundary conditions employed in simulations, for the comparison of our atomistic results with analytical theories. We analyse the induced charge distribution both in real and reciprocal space to quantify its lateral decay as a function of the Thomas-Fermi length and the distance of the ion from the first atomic plane of the electrode. We finally discuss how to bridge the atomistic and continuum pictures, which requires identifying the location of an effective sharp interface between the metal and vacuum.
Section~\ref{sec:systemmethod} introduces the considered system and its description by continuum electrostatics and atomistic simulations, while Section~\ref{sec:results} presents the results and discussion.


\section{System and methods}
\label{sec:systemmethod}

\subsection{Theory}
\label{sec:theory}

\begin{figure}[!ht]
\center
\includegraphics[width=4.5in]{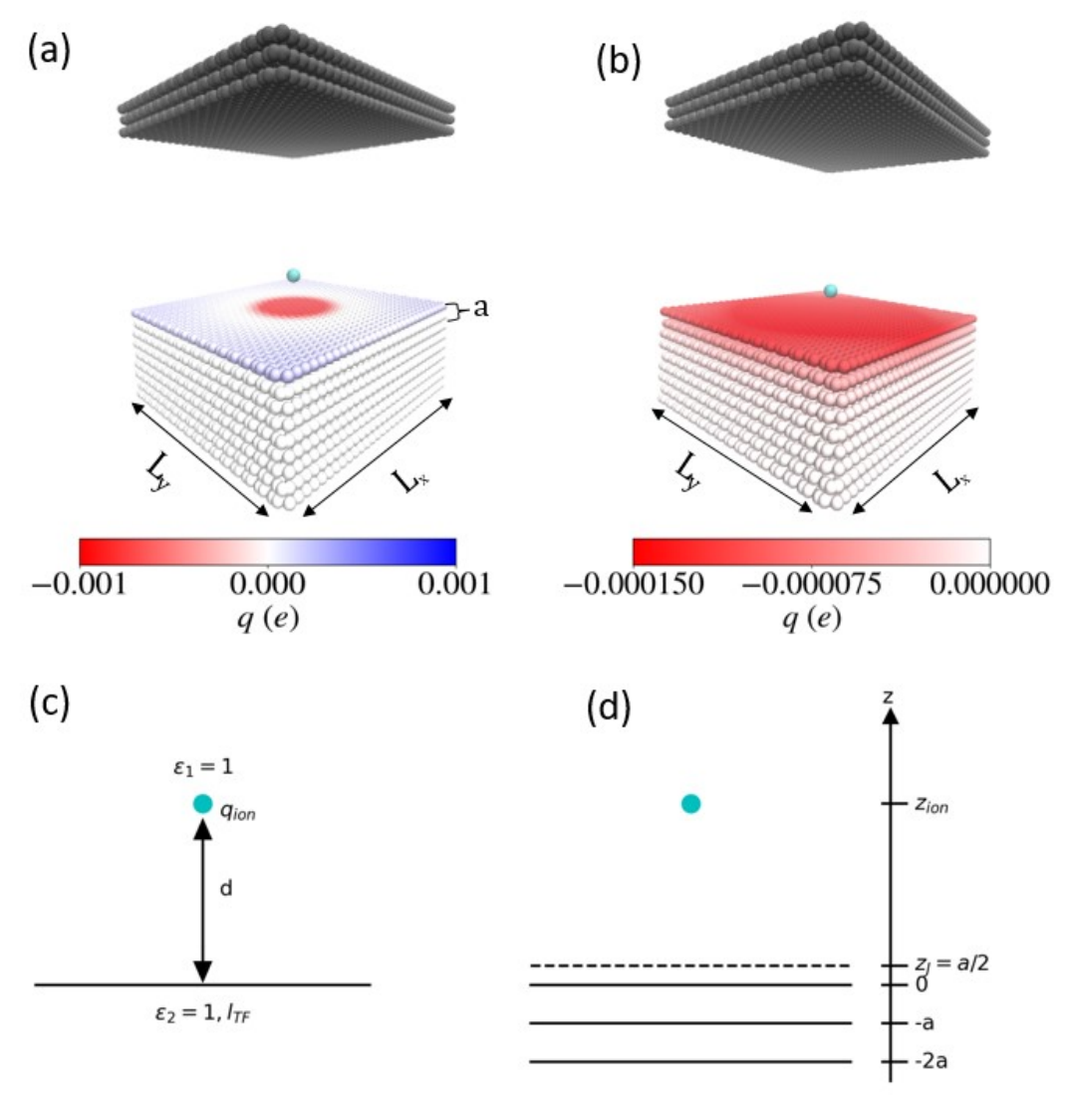}
\caption{
The system consists of a single ion in vacuum next to a Thomas-Fermi metal.
Panels (a) and (b) illustrate the corresponding simulation box, which contains a graphite electrode with explicit atoms (with interplane distance $a$), an ion (here located at a distance $z_{ion}=7.8$~\AA\ from the first atomic plane of the electrode, in cyan) and an insulating wall (in black); the color of electrode atoms indicates their charge induced by the ion, for a Thomas-Fermi screening length $l_{TF}=0$ and 5~\AA\, in panels (a) and (b), respectively. Periodic boundary conditions are used in the $x$ and $y$ directions only. 
Panel (c) shows the corresponding continuum picture for a single point charge $q_{ion}$ at a distance $d$ from the interface between two media: vacuum (with relative permittivity $\epsilon_1$) and a Thomas-Fermi metal, with a dielectric function $\epsilon(k)=\epsilon_2\times(1+k_{TF}^2/k^2)$ with $k_{TF}=l_{TF}^{-1}$ the reciprocal of the Thomas-Fermi screening length, $k$ the wavevector and $\epsilon_2=1$.
Panel (d) illustrates the issue of defining the location of the interface $z_{int}$ in the continuum picture from the discrete representation and the position $z_{ion}$ of the ion with respect to the first atomic plane; the position of the Jellium edge can be approximated as $z_J=+a/2$, with $a$ the interplane distance. 
}
\label{fig:snapshot}
\end{figure}

In the continuum picture, the charge induced in a Thomas-Fermi metal, characterized by a permittivity $\epsilon_2$ and a Thomas-Fermi screening length $l_{TF}$, by a point charge $q_{ion}$ in a polarizable medium characterized by a permittivity $\epsilon_1$  at a distance $d$ from the interface (see Fig.~\ref{fig:snapshot}c) can be computed from the electrostatic potential $\psi({\bf r})$. The latter satisfies $\Delta\psi_2=k_{TF}^2\psi_2$ inside the metal (defined here by the region $z<0$), with $k_{TF}=l_{TF}^{-1}$ the Thomas-Fermi wavevector, and $\Delta\psi_1=-\rho_{el}/\epsilon_0\epsilon_1$ outside ($z>0$), where $\rho_{el}({\bf r})=q_{ion}\delta({\bf r}-{\bf r}_{ion})$ is the volumic charge density and $\epsilon_0$ the vacuum permittivity, together with the continuity of the potential and of the electric displacement $-\epsilon\partial_z\psi$ at the interface. The solution to this problem can be found explicitly \emph{e.g.} in Ref.~\cite{Kaiser_2017}, using cylindrical coordinates (with $r$ the radial distance from the ion) and the Hankel transform of the observables such as the potential $f(z,k)=\int_0^\infty f(z,r) J_0(kr)r{\rm d}r$, with $J_0$ the zeroth order Bessel function of the first kind. The volumic induced charge density can then be computed as $\rho_{ind}^{3D}(z,k)=-\epsilon_0\epsilon_2k_{TF}^2\psi_2(z,k)$, with the result:
\begin{equation} \label{eq:hankel}
    {\rho}_{ind}^{3D}(z, K)=  \frac{-\epsilon_2 k_{\rm TF}^2 q_{ion} {e^{-kd + \sqrt{k^2 + k_{\rm TF}^2 } z}}} {2\pi({\epsilon_1 k + \epsilon_2 \sqrt{k^2 + k_{\rm TF}^2 )}}}
    \; .
\end{equation}
The inverse Hankel transform is not known analytically for finite $l_{TF}$, but this expression shows that the induced charge decays exponentially within the metal, and that the radial distribution depends on the depth. For a perfect metal ($l_{TF}=0$), the induced charge is localized at the interface ($z=0$) and the surface density is~\cite{Jackson}: 
\begin{equation} \label{eq:sigmaindltfzero}
    \sigma_{ind}^{l_{TF}=0}(r) = -\frac{q_{ion}}{2\pi}\frac{d}{(r^{2} + d^{2})^{3/2}}
\end{equation}

In the following, we will describe results from molecular simulations, in which periodic boundary conditions (PBC) in the directions parallel to the interface are employed. The external charge distribution inducing the polarization of the metal does not correspond to a single charge, but rather to a periodic array of charges $q_{ion}$ on a rectangular lattice, with dimensions $L_x$ and $L_y$. The solution of the corresponding electrostatic problem can be solved explicitly using cartesian coordinates as a sum over modes (see below). For finite $l_{TF}$, the induced charge also decays exponentially and the distribution in the $x$ and $y$ directions also depends on the depth $z$ within the metal. In order to analyze the effect of $l_{TF}$ on the lateral distribution, it is convenient to introduce the 2D charge density obtained by integrating over the $z$ direction inside the metal:
\begin{equation} \label{eq:2d}
    \sigma_{ind}(x,y) = \int_{-\infty}^0 \rho_{el}(x,y,z) \,{\rm d}z
    \; .
\end{equation}
The solution for finite $l_{TF}$, which can be obtained by considering the Fourier coefficients, reads:
\begin{equation} \label{eq:ana}
\sigma_{ind}(x,y) = \frac{-q_{ion}}{L_xL_y} \sum_{{m=0}}^{\infty}\sum_{{n=0}}^{\infty} a_{m,n}  \frac{\epsilon_2 k_{TF}^2 e^{-k_{m,n}d} \cos (k_m x) \cos (k_n y)}{\sqrt{k_{m,n}^2+k_{TF}^2}(\epsilon_1 k_{m,n} + \epsilon_2 \sqrt{k_{m,n}^2+k_{TF}^2})} 
\end{equation}
where we have introduced the wavevectors $k_m= m\times(2\pi/L_x)$, $k_n= n \times (2\pi/L_y)$ and $k_{m,n}=\sqrt{k_m^2 + k_n^2}$ as well as the weights
$a_{0,0}= 1$, $a_{m,0} = a_{0,n} = 2$ and $a_{m,n} = 4$.
While this result shares obvious similarities with the single charge case (see Eq.~\ref{eq:hankel} above for the 3D density, and Eq.~17 of Ref.~\cite{Kaiser_2017} for the 2D density), to the best of our knowledge the solution under PBC has not been reported previously. For comparison with the solution for a perfect metal Eq.~\ref{eq:sigmaindltfzero}, we will also compute the radial distribution:
\begin{equation} \label{eq:rad}
    \sigma_{ind}(r) = \frac{1}{2\pi r}\iint \sigma_{ind}(x,y)\delta\left(r-\sqrt{x^2 + y^2}\right) \,{\rm d}x{\rm d}y
    \; ,
\end{equation}
with $\delta$ Dirac's Delta function as well as its running integral:
\begin{equation} \label{eq:radialintegral}
    Q_{ind}(r) = \int_{0}^{r} \sigma_{ind}(r') 2\pi r' \,{\rm d}r'
    \; .
\end{equation}
Only $r<\min\{L_x,L_y\}/2$ is considered in the periodic case. In the present work, we consider the case of an interface between a Thomas-Fermi metal and vacuum, which corresponds to $\epsilon_1=\epsilon_2=1$.

\subsection{Simulation}

For an atomistic description of the metal polarization by an external charge, we use the constant-potential, fluctuating-charge model introduced by Siepmann and Sprik for the classical molecular simulation of water/electrode interfaces~\cite{Siepmann_1995}, later applied to electrochemical cells~\cite{reed2007a} and recently extended to take the Thomas-Fermi length into account~\cite{Scalfi_tfmodel_2020,scalfi-pnas-2021}. In this model, each electrode atom $i$ at position ${\bf r}_{i}$ with charge $q_{i}$ contributes to the total charge density as
\begin{equation} \label{eq:fl:q}
    \rho_{elec}({\bf r}) = \sum\limits_{i\in elec}\frac{q_i}{(2\pi w^2)^{3/2}}e^{-|{\bf r}-{\bf r}_i|^2/2w^2} \; ,
\end{equation} 
where $w$ is a fixed width of the Gaussian distribution around each atom. The instantaneous charges of all electrode atoms depend on the external charge distribution, \textit{i.e.} the positions ${\bf r}^N$ of atoms in the electrolyte (in the present study a single ion in vacuum). Specifically, they are determined by the constraints of constant potential imposed on each atom (one common value for all atom in a given electrode) and of global electroneutrality of the system~\cite{scalfi_charge_2020,scalfi_molecular_2021}. This process to compute the set of electrode charges ${\bf q}=\{q_i\}_{i\in elec}$ involves the potential energy of the system, which owing to the form of electrostatic interactions can be written as:
\begin{equation} \label{eq:utf}
    U({\bf {r}}^{N},{\bf{q}})= \frac{{\bf{q}}^{T} {\bf{A}}{\bf{q}}}{2} - {\bf{q}}^{T} {\bf{B}}({\bf {r}}^{N}) + C({\bf {r}}^{N})
    \; ,
\end{equation}
where the first term accounts for electrostatic interactions between electrode atoms, the second for those between electrode and electrolyte, and the third for all other interactions. The scalar $C$ is independent of electrode charges $\bf{q}$. Both the matrix ${\bf{A}}$ and vector $\bf{B}$ take into account Gaussian nature of the charge distribution of electrode atoms and the PBC inherent in molecular simulations of condensed matter systems. While this description was initially introduced for perfect metals, Scalfi \emph{et al.} showed that the Thomas-Fermi model can be recast in the same form by introducing an additional term in the matrix describing electrode-electrode interactions, namely~\cite{Scalfi_tfmodel_2020}:
\begin{equation}\label{eq:ltf}
    {\bf{A}}(l_{TF}) = {\bf{A}}_{0} + \frac{l_{TF}^{2} \bar{\rho}_{at}}{\epsilon_{0}} {\bf{I}}
    \; ,
\end{equation}
where $\bf{A_{0}}$ is the matrix for $l_{TF} = 0$, $\bar{\rho}_{at}$ is the atomic density of the electrode and $\bf {I}$ is the identity matrix. This description, already implemented in several simulation packages~\cite{marin-lafleche_metalwalls_2020,coretti_metalwalls_2022,ahrens-iwers_electrode_2022} allows to tune the metallic character of the metal without additional computational cost.

Specifically, we consider a system consisting of a graphite electrode, a single Na$^+$ ion in vacuum (with charge $q_{ion}=+e$, with $e$ the elementary charge), and an insulating neutral surface (which doesn't play a role in the present solvent-free case). This is illustrated in \fref{fig:snapshot}(a) and (b), with the color indicating the charge of each electrode atom for $l_{TF}=0$ and 5~\AA, respectively. The graphite electrode, maintained at a fixed potential (whose value is irrelevant since there is no counter-electrode), consists of 17280 atoms in an hexagonal lattice arranged in 9 atomic planes perpendicular to the $z$ direction. The distance between carbon atoms within a plane is $d_{CC}=1.42$~\AA, while the distance between planes is $a=3.354$~\AA. The width of the Gaussian distribution around each atom $w= 0.4$~\AA, and the Thomas-Fermi length $l_{TF}$ ranges from $0$~\AA\, to $5$~\AA. Even for the largest $l_{TF}$ considered, the electrode is sufficiently thick to ensure that the charges on the last atomic planes are negligible. The insulating wall consists of 5760 carbon atoms arranged in 3 planes with the same graphite structure. The atomic plane closest to the ion is located at $z=0$ for the metal, while that of the insulating wall is at $z=55.088$~\AA. In order to investigate the effect of the ion-electrode distance on the induced charge distribution, we consider positions $z_{ion} = 5.1, 7.8, 11.0 \,\text{and}\, 15.0$~\AA. These values were chosen based on the equilibrium density profile of Na$^{+}$ at the interface between graphite and a 1~M NaCl aqueous solution using the same model (see Appendix). The box dimensions are $L_x = 68.20$~\AA, $L_y = 73.83$~\AA\, and $L_z = 231.79$~\AA, and PBC are applied only in the $x$ and $y$ directions. 

Electrostatic interactions are computed by 2D Ewald summation, taking into account the Gaussian distributions of electrode atoms and the point charge of the ion~\cite{coretti_metalwalls_2022}. Even though they do not play a role in the present calculation, we also introduce non-electrostatic interactions via truncated and shifted Lennard-Jones (LJ) potentials, using the LJ parameters for Na$^+$ and carbon from Refs.~\cite{Dang-jacs-1995} and~\cite{Werder-jpcb-2003}, respectively, with Lorentz-Berthelot mixing rules. All simulations are performed using the molecular dynamics code Metalwalls~\cite{marin-lafleche_metalwalls_2020}, and the charges on the electrode atoms are determined using matrix inversion~\cite{scalfi_charge_2020}. Since there is no solvent and the ion is fixed, a single simulation step is sufficient to determine the induced charge on each atom. The 3D charge density can then be reconstructed with Eq.~\ref{eq:fl:q} and integrated as indicated in Eq.~\ref{eq:2d} to obtain the 2D charge density. In practice, it is more efficient to perform the integration over $z$ analytically and simply reconstruct the 2D distribution directly as
\begin{equation} \label{eq:2d:MD}
    \sigma_{ind}^{MD}(x,y) = \sum\limits_{i\in elec}\frac{q_i}{2\pi w^2}e^{-[(x-x_i)^2+(y-y_i)^2]/2w^2} \; .
\end{equation} 
This quantity is computed on a 2D grid with bin width $0.1$~\AA. The radial distribution can then be obtained numerically by Eq.~\ref{eq:rad}.

\section{Results and Discussion}
\label{sec:results}

In the following, we present the results for the charge induced on the electrode obtained with the atomistic description, compare them with various analytical predictions within the continuum picture and discuss in particular the links between the two levels of description. The snapshots introducing the system in Fig.~\ref{fig:snapshot} illustrate the main effects of introducing a finite Thomas-Fermi screening length. While for a perfect metal ($l_{TF}=0$~\AA, panel~\ref{fig:snapshot}a) the charge is essentially localized in the first atomic plane of the electrode and decays away from the atom, for $l_{TF}=5$~\AA\, a significant charge is induced on at least one more atomic plane and the lateral distribution visible in the first plane is much more homogeneous. The exponential decay of the charge per plane within the electrode (with a characteristic length $l_{TF}$) was demonstrated for capacitors with vacuum or an electrolyte solution between two electrodes in Ref.~\cite{Scalfi_tfmodel_2020} introducing the Thomas-Fermi model in constant-potential, fluctuating charge simulations. Here we focus on the lateral distribution of the induced charge, by considering the integrated charge density $\sigma_{ind}(x,y)$ defined by Eq.~\ref{eq:2d}. We begin with a qualitative analysis of 2D maps of the induced charge in Section~\ref{sec:results:2Dmaps}, before turning to a quantitative analysis of the radial charge distribution in Section~\ref{sec:results:radial} and of the 2D distribution in reciprocal space in Section~\ref{sec:results:reciprocal}.

\subsection{Qualitative analysis with 2D maps of the induced charge}
\label{sec:results:2Dmaps}

\subsubsection{Effect of ion-surface distance for a perfect metal}
\label{sec:results:2Dmaps:dion}

\begin{figure}[!ht]
\center
\includegraphics[width=5.5in]{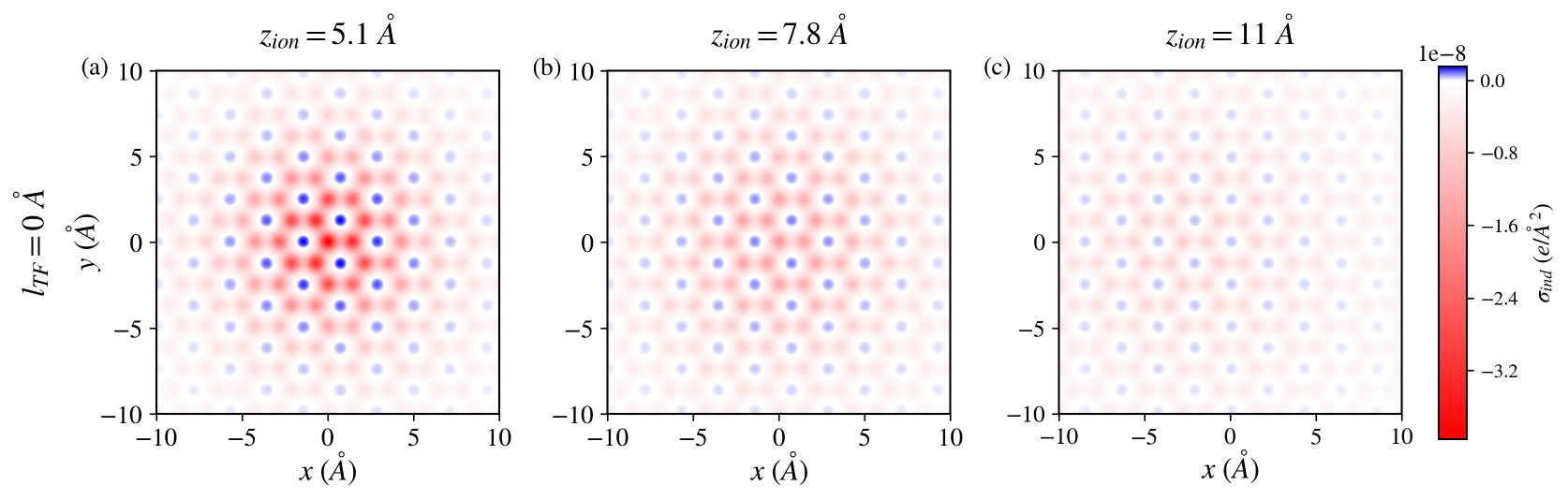}
\caption{
Surface charge density $\sigma_{ind}(x,y)$ (see Eq.~\ref{eq:2d:MD}) induced in the graphite electrode, described as a perfect metal ($l_{TF}=0$~\AA), by an ion located at $(x,y)=(0,0)$ and a distance from the first atomic plane of the electrode $z_{ion}=5.1, 7.8$ and $11$~\AA\, for panels (a), (b) and (c), respectively. The colormap is common for the 3 panels. The slightly positively charged regions (blue) correspond to atoms in the second atomic plane of the electrode.}
\label{fig:2d_map_d}
\end{figure}

We start by briefly recalling the effect of an ion-surface distance for the case of a perfect metal ($l_{TF}=0$), already investigated in Ref.~\cite{pireddu_molecular_2021} for a gold electrode (FCC lattice structure) and a slightly different setup including a countercharge on the insulating wall (resulting with the ion in a net zero charge induced on the electrode). Fig.~\ref{fig:2d_map_d} shows the surface charge density $\sigma_{ind}(x,y)$ induced in the graphite electrode, described as a perfect metal ($l_{TF}=0$~\AA), by a Na$^+$ ion located at $(x,y)=(0,0)$ and a distance from the first atomic plane of the electrode $z_{ion}=5.1, 7.8$ and $11$~\AA\, for panels (a), (b) and (c), respectively. 

Despite an obvious difference, namely the spatial variations associated with the discrete atomic structure of the electrode (modulated by the Gaussian distribution around each atom, see also Section~\ref{sec:results:reciprocal}), the results are overall consistent with the continuum picture. For each distance of the Na$^+$ ion from the surface, the induced charge is more negative close to the ion and decays away from it, consistently with Eq.~\ref{eq:sigmaindltfzero}. The slightly positively charged regions (blue) correspond to atoms in the second atomic plane of the electrode and likely results from the nearby large negative charges, as already observed in Ref.~\cite{pireddu_molecular_2021}. Increasing the distance of the ion from the surface results in a less negative charge density close to the ion and in a slower lateral decay, consistently with Eq.~\ref{eq:sigmaindltfzero}, and in a more homogeneous charge distribution (see also Section~\ref{sec:results:radial:pbc}).

\subsubsection{Effect of the Thomas-Fermi screening length}
\label{sec:results:2Dmaps:ltf}

\begin{figure}[!ht]
\center
\includegraphics[width=5.5in]{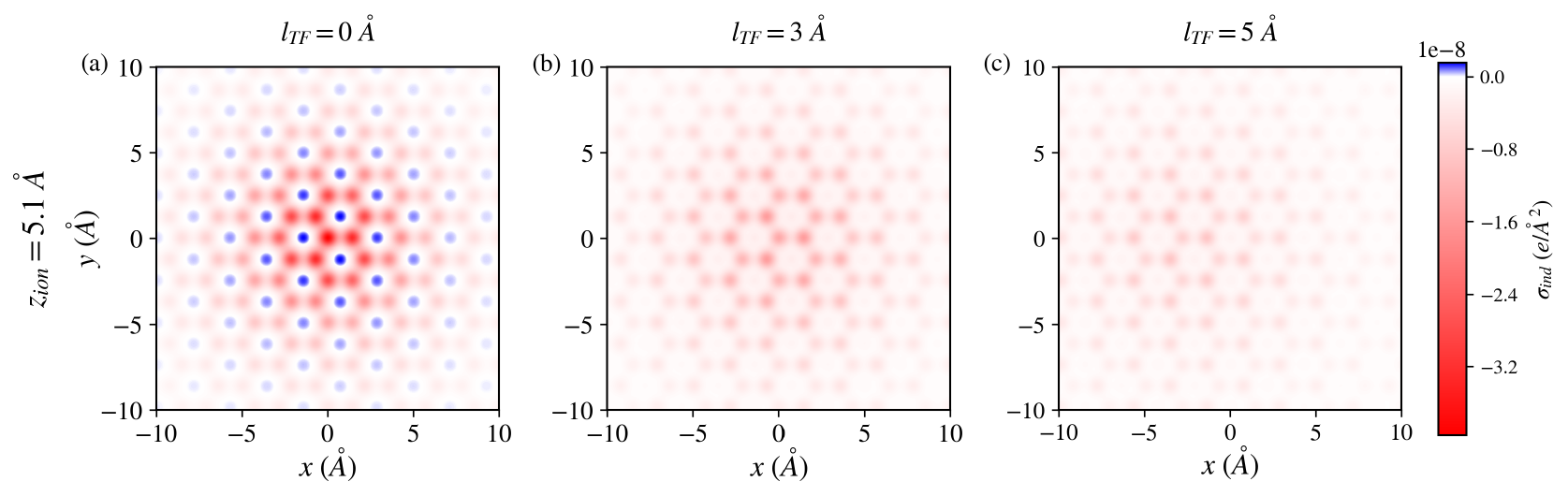}
\caption{
Surface charge density $\sigma_{ind}(x,y)$ (see Eq.~\ref{eq:2d:MD}) induced in the graphite electrode by an ion located at $(x,y)=(0,0)$ and a distance $z_{ion}=5.1$~\AA\, from the first atomic plane of the electrode, for Thomas-Fermi lengths $l_{TF}=0, 3$ and $5$~\AA\, in panels (a), (b) and (c), respectively. The colormap is common for the 3 panels and the same as in Fig.~\ref{fig:2d_map_d}.}
\label{fig:2d_map_ltf}
\end{figure}

We now turn to the effect of the Thomas-Fermi screening length on the charge distribution induced by the ion, which had not been considered previously. Fig.~\ref{fig:2d_map_ltf} shows the surface charge density $\sigma_{ind}(x,y)$ for an ion located at $(x,y)=(0,0)$ and a distance $z_{ion}=5.1$~\AA\, from the first atomic plane of the electrode, for $l_{TF}=0, 3$ and $5$~\AA\, in panels (a), (b) and (c), respectively (panel~\ref{fig:2d_map_ltf}a is therefore identical to panel~\ref{fig:2d_map_d}a). Consistently with the continuum picture, which will be discussed in more detail below, the lateral decay of the charge is slower as $l_{TF}$ increases, resulting in a more homogeneous charge distribution. This is also in line with the increase with $l_{TF}$ of the correlation length observed in the surface charge induced in a capacitor consisting in an aqueous NaCl solution between graphite electrodes~\cite{scalfi-pnas-2021}. Such charge correlations within the electrode can be understood as resulting from the energetic cost of localizing the charge on each atom, evident in the second term in the r.h.s. Eq.~\ref{eq:ltf}, which entails an energy penalty $\propto l_{TF}^2 \sum_i q_i^2$ in Eq.~\ref{eq:fl:q}. This penalization of large local charges also results in the absence of positively charged regions in panels~\ref{fig:2d_map_ltf}b and~\ref{fig:2d_map_ltf}c, which originated in the second plane due to the high local charge in the first plane for $l_{TF}=0$.

\subsection{Quantitative analysis with radial distribution of the induced charge}
\label{sec:results:radial}

\subsubsection{Effect of periodic boundary conditions}
\label{sec:results:radial:pbc}

For the quantitative comparison of the results from atomistic calculations with analytical predictions, it is important to acknowledge the effect of periodic boundary conditions. This effect was already emphasized in Ref.~\cite{pireddu_molecular_2021} for gold electrodes with $l_{TF}=0$, where it was taken into account by summing over periodic images their contribution in real space, Eq.~\ref{eq:sigmaindltfzero}. Since we don't know analytically the corresponding real-space expression for finite $l_{TF}$, in the present work we use instead the solution for the periodic case expressed as a sum over modes, Eq.~\ref{eq:ana}. We illustrate the effect of PBCs in the $l_{TF}=0$ case, for which the periodic solution can be expressed as the $k_{TF}\to\infty$ limit of Eq.~\ref{eq:ana}:
\begin{equation} \label{eq:ana:pbc:ltfzero}
\sigma_{ind}^{l_{TF}=0}(x,y) = \frac{-q_{ion}}{L_xL_y} \sum_{{m=0}}^{\infty}\sum_{{n=0}}^{\infty} a_{m,n}  e^{-k_{m,n}d} \cos (k_m x) \cos (k_n y)
\end{equation}
with the same definition of the wavevectors $k_m, k_n$ and $k_{m,n}$ and weights $a_{m,n}$. It is possible to show that in the limit $L_x,L_y\to\infty$, replacing the discrete sum by an integral and using polar coordinates, one recovers the solution for a single charge Eq.~\ref{eq:sigmaindltfzero} with $r=\sqrt{x^2+y^2}$.

\begin{figure}[!ht]
\center
\includegraphics[width=3in]{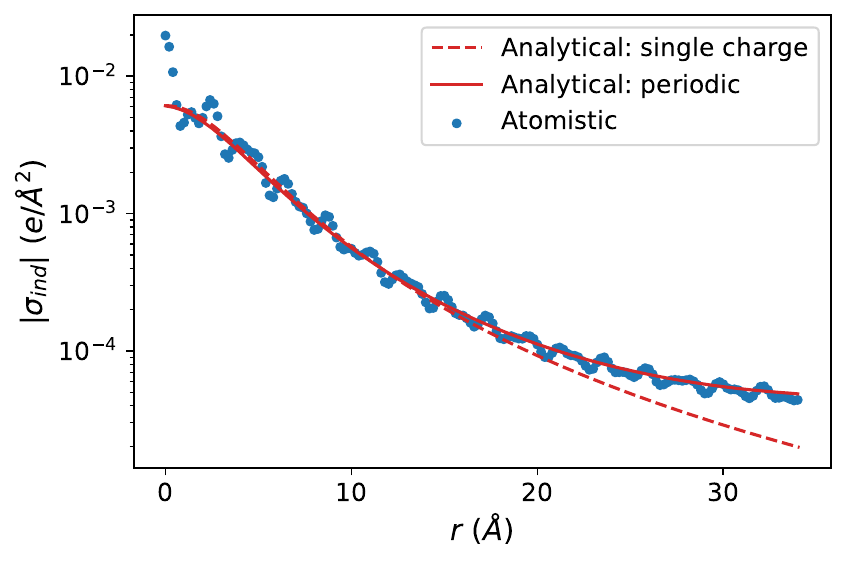}
\caption{
Radially averaged surface charge density $\sigma_{ind}(r)$ (see Eq.~\ref{eq:rad}) induced in the graphite electrode, described as a perfect metal ($l_{TF}=0$~\AA), by an ion located at $(x,y)=(0,0)$ and a distance from the first atomic plane of the electrode $z_{ion}=5.1$~\AA. The results obtained within the atomistic description (symbols) are compared with the analytical prediction for a single charge (Eq.~\ref{eq:sigmaindltfzero}, dashed line) and that taking into account the periodic boundary conditions via Eq.~\ref{eq:ana} (solid line). 
}
\label{fig:pbc}
\end{figure}

The analytical predictions for both the single charge and periodic cases correspond to Eqs.~\ref{eq:sigmaindltfzero} and~\ref{eq:ana:pbc:ltfzero} require the knowledge of the distance $d$ of the ion from the interface. This is an important issue, discussed in more detailed below. Here we simply use $d=z_{ion}$, \textit{i.e.} the distance of the ion from the first atomic plane of the electrode. As shown in Fig.~\ref{fig:pbc} the radially averaged surface charge density $\sigma_{ind}(r)$ for $l_{TF}=0$ and $z_{ion}=5.1$~\AA\, obtained from the atomistic description is well described (except for the oscillations due to the graphite lattice) by the analytical solution Eq.~\ref{eq:ana:pbc:ltfzero} for the periodic case over the whole $r$ range. Expectedly, this is not the case for the single-charge result Eq.~\ref{eq:ana} which deviates from the numerical results at large distance. In the following, we therefore only consider the periodic solution Eq.~\ref{eq:ana} (and Eq.~\ref{eq:ana:pbc:ltfzero} for $l_{TF}=0$).

\subsubsection{Effect of ion-surface distance for a perfect metal}
\label{sec:results:radial:dion}

\begin{figure}[!ht]
\center
\includegraphics[width=5.5in]{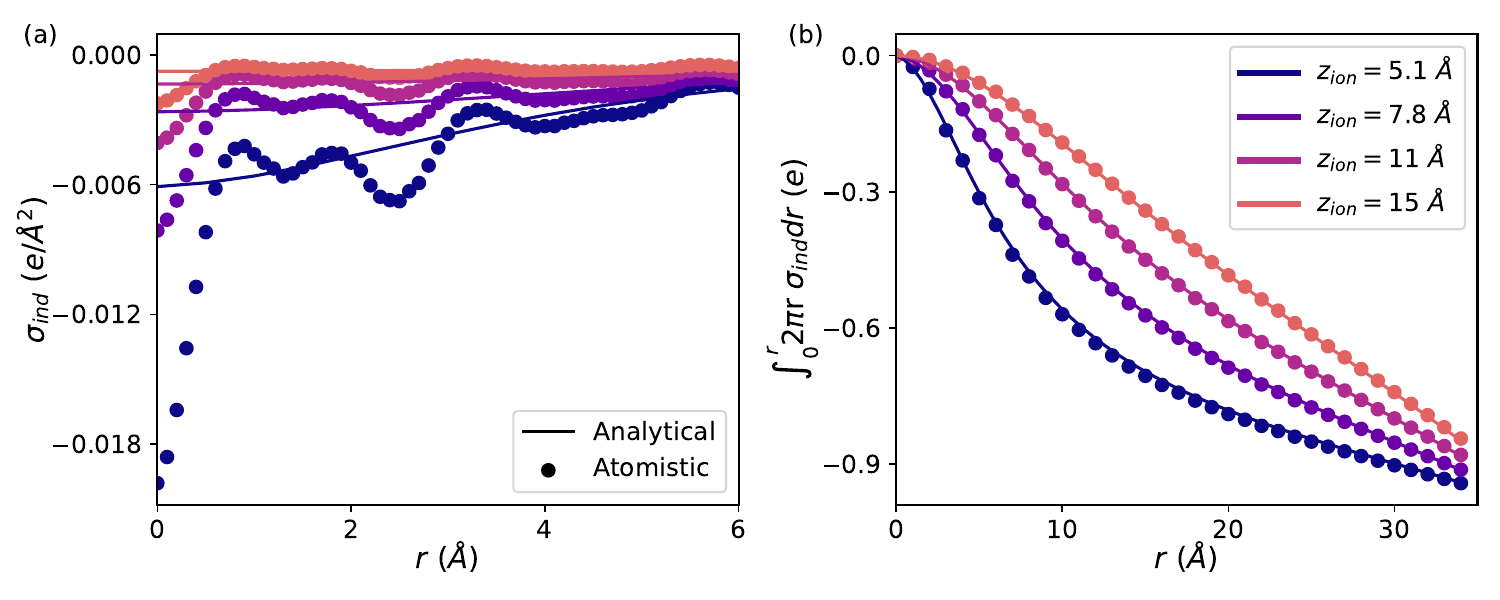}
\caption{
(a) Radially averaged surface charge density $\sigma_{ind}(r)$ (see Eq.~\ref{eq:rad}) induced in the graphite electrode, described as a perfect metal ($l_{TF}=0$~\AA), by an ion located at $(x,y)=(0,0)$ and a distance from the first atomic plane of the electrode $z_{ion}=5.1, 7.8, 11$ and $15$~\AA\, indicated by the colors in panel (b). The results obtained within the atomistic description (symbols) are compared with the analytical prediction taking into account the periodic boundary conditions via Eq.~\ref{eq:ana} (solid lines). 
(b) Running integral of the charge densities shown in the previous panel (see Eq.~\ref{eq:radialintegral}).
}
\label{fig:radial:ltfzero}
\end{figure}

As evident from Eq.~\ref{eq:sigmaindltfzero}, the induced charge distribution is expected to depend on the distance of the ion from the interface. Fig.~\ref{fig:radial:ltfzero} reports the atomistic prediction for $\sigma_{ind}(r)$ for $l_{TF}=0$ and various positions $z_{ion}$ of the ion, as well as the corresponding radial integrals $Q_{ind}(r)$ (see Eq.~\ref{eq:radialintegral}) in panels (a) and (b), respectively. In both panels, we also indicate the analytical predictions  Eq.~\ref{eq:sigmaindltfzero} using $d=z_{ion}$ as the distance from the interface. As already reported for a gold electrode in Ref.~\cite{pireddu_molecular_2021}, Eq.~\ref{eq:sigmaindltfzero} expectedly fails to capture the oscillations of $\sigma_{ind}(r)$, which are due to the atomic structure of the electrode (this point will be further examined in Section~\ref{sec:results:reciprocal}). These oscillations are averaged out in the radial integral, which is very well described by the analytical prediction from continuum electrostatics. Increasing the distance of ion from the surface decreases the local charge below the ion ($r=0$) and increases the lateral extent of the charge distribution (with characteristic length $d$ in the non-periodic case).

\subsubsection{Effect of the Thomas-Fermi screening length}
\label{sec:results:radial:ltf}

\begin{figure}[!ht]
\center
\includegraphics[width=5.5in]{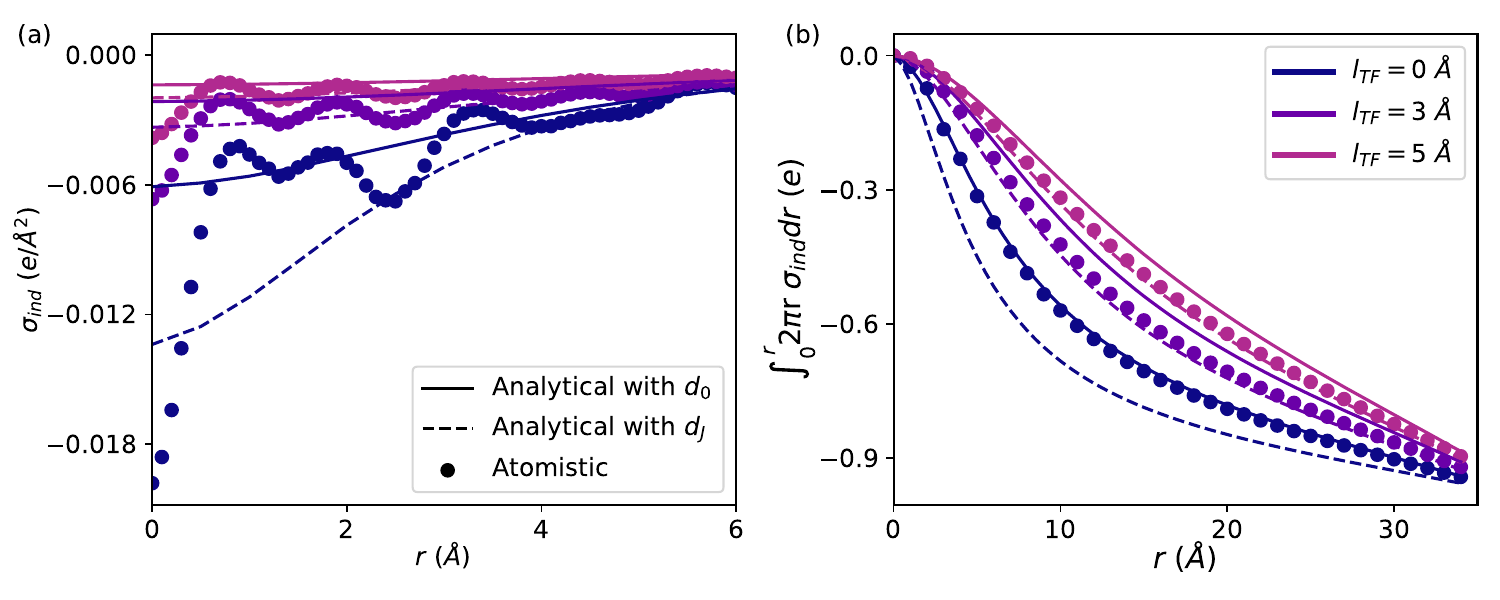}
\caption{
(a) Radially averaged surface charge density $\sigma_{ind}(r)$ (see Eq.~\ref{eq:rad}) induced in the graphite electrode by an ion located at $(x,y)=(0,0)$ and a distance $z_{ion}=5.1$~\AA\, from the first atomic plane of the electrode, for Thomas-Fermi lengths $l_{TF}=0, 3$ and $5$~\AA\, indicated by the colors in panel (b). The results obtained within the atomistic description (MD, symbols) are compared with the analytical prediction taking into account the periodic boundary conditions via Eq.~\ref{eq:ana}, considering the position of the interface to be located at the first atomic plane (solid lines) or at the Jellium edge (dashed lines), see Eq.~\ref{eq:dzerodj} and panel~\ref{fig:snapshot}d. 
(b) Running integral of the charge densities shown in the previous panel (see Eq.~\ref{eq:radialintegral}).
}
\label{fig:radial:ltf}
\end{figure}

We now turn to the effect, which was not considered previously, of the Thomas-Fermi screening length on $\sigma_{ind}(r)$. Fig.~\ref{fig:radial:ltf} reports the atomistic prediction for $z_{ion}=5.1$~\AA\, and three values of $l_{TF}$ (0, 3 and 5~\AA), as well as the corresponding radial integrals $Q_{ind}(r)$ (see Eq.~\ref{eq:radialintegral}), in panels (a) and (b), respectively. Qualitatively, comparison with Fig.~\ref{fig:radial:ltfzero} suggests that the effect of increasing $l_{TF}$ is similar to that of increasing the distance $d$ from the ion, with a decrease of the local charge below the ion ($r=0$) and an increase of the lateral extent of the charge distribution. 

For a quantitative analysis, both panels in Fig.~\ref{fig:radial:ltf} also report the results from the continuum prediction, namely for finite $l_{TF}$ the radial average Eq.~\ref{eq:rad} of Eq.~\ref{eq:ana} and the corresponding running integral Eq.~\ref{eq:radialintegral}. As in the $l_{TF}=0$ case, the continuum prediction fails to account for the oscillations due to the atomic lattice. In addition, the predictions using $d=z_{ion}$ as the distance from the interface, which quantitatively agrees with the atomistic result for $l_{TF}=0$, underestimates the magnitude of the local charge close to the ion and overestimates the extent of the lateral distribution. 

Scalfi \emph{et al.} showed in Ref.~\cite{Scalfi_tfmodel_2020} that in order to quantitatively describe the capacitance per unit area of empty capacitors by $\epsilon_0/L_{vac}$, with $L_{vac}$ the width of the vacuum slab between the electrodes, for finite $l_{TF}$ one must consider that the interface between the metal and vacuum (the so-called ``Jellium edge'') is located away from the first atomic plane of the electrode, at a distance $z_J=a/2$, with $a$ the interplane distance within the electrode (see Fig.~\ref{fig:snapshot}d). One can therefore distinguish two predictions corresponding to the same model with two definitions of the position of the interface, $z_{int}^0=0$ and $z_{int}^J=z_J$ corresponding to the first atomic plane and to the Jellium edge, respectively, leading to different definitions of the ion-surface distance:
\begin{equation}\label{eq:dzerodj}
    \left\{
    \begin{array}{ll}
        d_{0} &= z_{ion} - z_{int}^{0} = z_{ion} \vspace{0.2cm}\\ 
        d_{J} &= z_{ion} - z_{int}^{J} = z_{ion} - a/2
    \end{array}
    \right.
\end{equation}
Both panels of Fig.~\ref{fig:radial:ltf} include the analytical predictions using Eq.~\ref{eq:ana} with $d_0$ and $d_J$ as solid and dashed lines, respectively. The definition $d_J$ provides a better agreement with the numerical results, in particular for the radial integral in panel~\ref{fig:radial:ltf}b, when the Thomas-Fermi length is comparable to or larger than the distances between atoms ($d_{CC}$ within planes and $a$ between planes), while $d_0$ is more accurate for $l_{TF}=0$ ($d_J$ overestimates the local charge and underestimates the lateral extent in that case). The definition of the interface position will be further discussed in Section~\ref{sec:results:reciprocal:ltf}.

\subsection{Quantitative analysis in reciprocal space}
\label{sec:results:reciprocal}

\subsubsection{Definition of a characteristic decay length}
\label{sec:results:reciprocal:def}

For a single charge, $\sigma_{ind}(r)$ decays to 0 sufficiently fast that a characteristic length for the lateral decay of the induced charge can be defined \emph{e.g.} as:
\begin{equation}
    \xi = \frac{\int_0^{\infty} r\, \sigma_{ind}(r) 2 \pi r \, {\rm d}r}{\int_0^{\infty} \sigma_{ind}(r) 2 \pi r \, {\rm d}r}
\end{equation}
which simplifies using Eq.~\ref{eq:sigmaindltfzero} to $\xi=d$ in the perfect metal case $l_{TF}=0$. However, such a definition cannot be used for periodic systems, where the reciprocal space provides a more convenient way to express and analyze the data, as evident from the expression of the induced charge Eq.~\ref{eq:ana} as a sum over modes. In fact, Eq.~\ref{eq:ana:pbc:ltfzero} for $l_{TF}=0$ can also be rewritten as a sum over periodic images of the contributions of the form Eq.~\ref{eq:sigmaindltfzero} (this is how the periodic solution was computed numerically in Ref~\cite{pireddu_molecular_2021}). The presence of the exponential term $e^{-k_{m,n}d}$ results, in the limit $L_x,L_y\to\infty$ where the discrete sum can be replaced by an integral, in Eq.~\ref{eq:sigmaindltfzero} for a single charge, because $d/(r^2+d^2)^{3/2}$ is the inverse Hankel transform of $e^{-kd}$. The characteristic decay length can therefore also be defined from the behaviour in reciprocal space.

More precisely, and in line with the analysis of Vorotyntsev and Kornyshev~\cite{vorotyntsev_1980} from the behaviour at large $r$, for finite $l_{TF}$ we consider the corresponding $k\to0$ behaviour of the induced surface charge density. Starting from Eq.~\ref{eq:ana}, and assuming $k\ll \epsilon_1 k_{TF}/\epsilon_2$ (and $k\ll k_{TF}$, which is implied by the previous relation in the general case where $\epsilon_2\geq\epsilon_1$) we approximate the prefactor of each mode as:
\begin{align} \label{eq:anaapprox}
\sigma_{ind}(x,y) &\approx \frac{-q_{ion}}{L_xL_y} \sum_{{m=0}}^{\infty}\sum_{{n=0}}^{\infty} a_{m,n}  e^{-k_{m,n}d} \left(1-\frac{ \epsilon_1 k_{m,n}}{\epsilon_2 k_{TF}} \right) \cos (k_m x) \cos (k_n y) 
\nonumber \\
&\approx \frac{-q_{ion}}{L_xL_y} \sum_{{m=0}}^{\infty}\sum_{{n=0}}^{\infty} a_{m,n}  e^{-k_{m,n}d_{eff}} \cos (k_m x) \cos (k_n y) 
\end{align}
with an effective distance
\begin{equation} \label{eq:deff}
    d_{eff} = d + \frac{\epsilon_1 l_{TF}}{\epsilon_2}
    \; ,
\end{equation}
which in the present case where $\epsilon_1=\epsilon_2$ simplifies to $d_{eff}=d+l_{TF}$. In the limit $L_x,L_y\to\infty$, this leads to approximating the exact solution for a single charge and finite $l_{TF}$ by the solution for $l_{TF}=0$, Eq.~\ref{eq:sigmaindltfzero}, with an effective distance $d_{eff}$. This approximation, which is expected to be more accurate at large than short distance $r$, is consistent with the qualitative observations of Section~\ref{sec:results:radial:ltf}.

In order to examine the charge density in reciprocal space, we consider the Fourier coefficients:
\begin{equation} \label{eq:tildesigma}
    \tilde{\sigma}_{ind}(k_m,k_n) = \frac{1}{L_x L_y} \int_0^{L_x} \int_0^{L_y}\sigma_{ind}(x,y) e^{i(k_m x + k_n y)} \, {\rm d}x {\rm d}y
\end{equation}
where the wavevectors are the ones compatible with the periodicity of the system, defined below Eq.~\ref{eq:ana}. In the continuum picture, the latter equation yields
\begin{equation} \label{eq:tildesigma:ltf}
    \tilde{\sigma}_{ind}(k_m,k_n) = -\frac{q_{ion}}{L_xL_y}\frac{\epsilon_2 k_{TF}^2 e^{-k_{m,n}d} }{\sqrt{k_{m,n}^2+k_{TF}^2}(\epsilon_1 k_{m,n} + \epsilon_2 \sqrt{k_{m,n}^2+k_{TF}^2})} 
\end{equation}
which reduces to
\begin{equation} \label{eq:tildesigma:ltfzero}
    \tilde{\sigma}_{ind}^{l_{TF}=0}(k_m,k_n) = -\frac{q_{ion}}{L_xL_y}e^{-k_{m,n}d}
\end{equation}
for $l_{TF}=0$. In the $k\to0$ limit, Eq.~\ref{eq:tildesigma:ltf} can be approximated by 
\begin{equation} \label{eq:tildesigma:kzero}
    \tilde{\sigma}_{ind}^{l_{TF}}(k_m,k_n) \approx -\frac{q_{ion}}{L_xL_y}e^{-k_{m,n}d_{eff}}
    \, ,
\end{equation}
with $d_{eff}$ given by Eq.~\ref{eq:deff}. For the atomistic results, the Fourier components can be computed numerically via 2D Fast Fourier Transform (FFT). In the continuum picture the induced charge is real and even with respect to both $x$ and $y$, so that the Fourier coefficients are real and even with respect to $k_m$ and $k_n$. In the atomistic case, the structure of the graphite lattice (see inset in panel~\ref{fig:fourier}b) results in both an even real part and an odd imaginary part. In the following, in order to compare the two levels of description we only report the real part, and further limit the range of wavevectors to  $k_{m,n}\geq0$.

\subsubsection{2D charge density in reciprocal space: perfect metal}
\label{sec:results:reciprocal:ltfzero}

\begin{figure}[!ht]
\center
\includegraphics[width=5.5in]{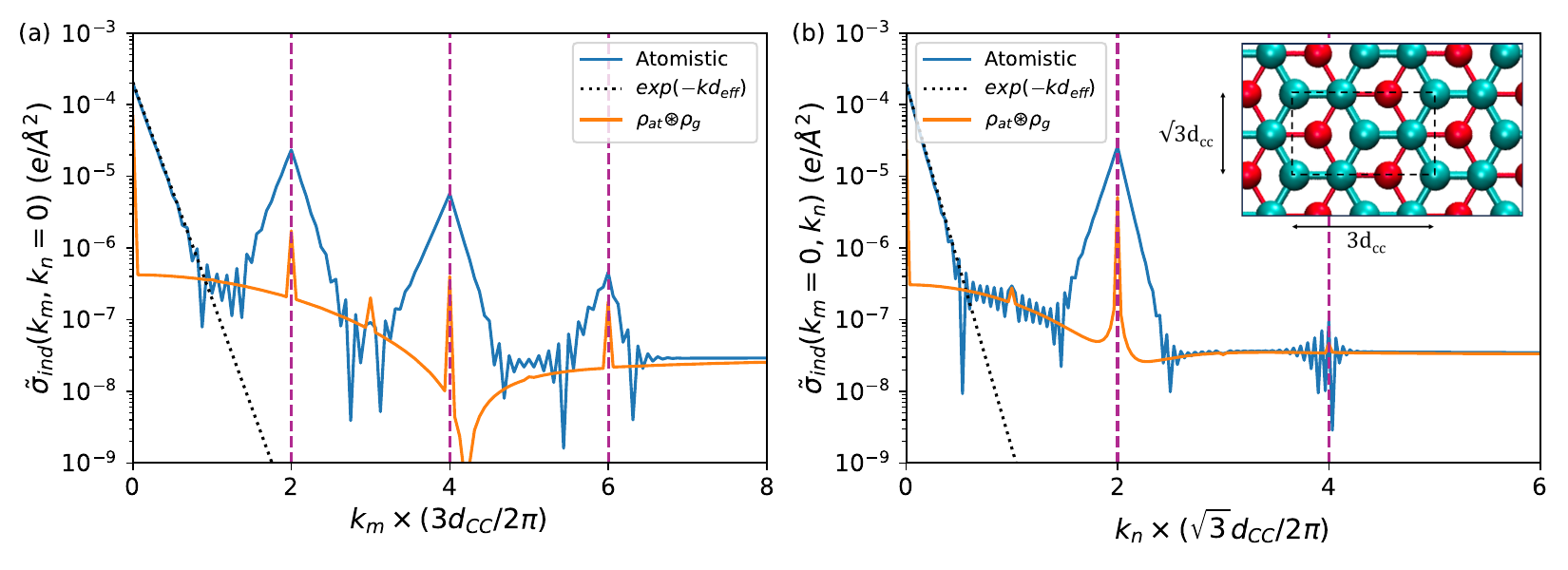}
\caption{
Real part of the Fourier coefficients of the two-dimensional induced charge density, defined in Eq.~\ref{eq:tildesigma}, for $z_{ion}=5.1$~\AA\, and $l_{TF}=0$~\AA. Panel (a) corresponds to $\tilde{\sigma}_{ind}(k_m,k_n=0)$, while panel (b) corresponds to $\tilde{\sigma}_{ind}(k_m=0,k_n)$. In each panel, the wavevector is rescaled by $2\pi/b$, with $b$ the relevant dimension of the unit cell of the graphite lattice. The latter is shown as dashed lines in the inset of panel (b), where cyan and red atoms correspond to the first and second atomic planes, respectively (half of the atoms in the second plane are hidden below atoms in the first plane).
In both panels, the results from the atomistic description (blue solid line) are compared with the continuum prediction Eq.~\ref{eq:tildesigma:ltfzero} with $d$ fitted in the $k\to0$ region to extract an effective distance $d_{eff}$ (dotted lines). We also show (orange line) the Fourier coefficients of the convolution of the atomic number density $\rho_{at}$ with the Gaussian charge density $\rho_g$ for each atom (see text). The vertical dashed lines indicate even values of the abscissa.
}
\label{fig:fourier}
\end{figure}

Fig.~\ref{fig:fourier} shows the real part of the Fourier coefficients for $z_{ion}=5.1$~\AA\, and $l_{TF}=0$~\AA, for $k_n=0$ and $k_m=0$ in panels~\ref{fig:fourier}a and~\ref{fig:fourier}b, respectively. In each panel, the wavevector is rescaled by $2\pi/b$, with $b$ the relevant dimension of the unit cell of the graphite lattice, shown as dashed lines in the inset of panel~\ref{fig:fourier}b. In both directions, one observes three features: (1) peaks at even multiples of $2\pi/b$, where $b=3d_{CC}$ and $\sqrt{3}d_{CC}$ in the $x$ and $y$ directions, respectively; (2) wings decaying exponentially around each peak, appearing as straight lines on the semi-log plot (dotted lines in Fig.~\ref{fig:fourier}); (3) a baseline between the peaks.   

The peaks reflect the discrete lattice structure, while the baseline is due to the Gaussian distribution of the charge around each atom. This can be confirmed by considering the density obtained by considering the atomic density $\rho_{at}$ convoluted by the same Gaussian distribution $\rho_g$, \emph{i.e} setting $q_i=1$ for all atoms in Eq.~\ref{eq:2d:MD}. We estimate $\rho_{at}$ numerically by summing over atoms in the first two planes (since two planes are needed to include the full graphite unit cell); to facilitate comparison with the charge distribution in the figure, we divide the result by the number of atoms in the two planes (since for $k=0$ we expect $q_{ion}/L_xL_y$ for the charge density and $N_{at}/L_xL_y$ for the number density). Some $(x_i,y_i)$ coordinates correspond to atoms in both atomic planes of the graphite unit cell (see the inset in panel~\ref{fig:fourier}b). The Fourier coefficients of the convolution $\rho_{at} \circledast \rho_g$, shown as orange solid line in Fig.~\ref{fig:fourier}, reproduce the location of the peaks and the baseline. The magnitude of the peaks differs however from that of $\tilde{\sigma}_{ind}$ because the charge distribution is not identical to the atomic distribution.

Finally, in the $k\to0$ limit the Fourier coefficients decay exponentially, consistently with Eq.~\ref{eq:tildesigma:ltfzero}, or more generally with Eq.~\ref{eq:anaapprox}. The slopes of the decay in the $x$ and $y$ directions result in the same effective decay lengths $d_{eff}=4.8\pm0.1$~\AA. This value is consistent with the distance $d_0=5.1$~\AA\, for this case, even though slightly smaller. Possible deviations could be due to the breakdown of the continuum approximation for this relatively short distance compared to the molecular features of graphite ($d_{CC}$ and corresponding unit cell parameters in the plane, $a$ between planes), or to the Gaussian charge distribution on electrode atoms, as already noted for the effective distance necessary between electrodes to reproduce the capacitance of empty capacitors~\cite{Scalfi_tfmodel_2020,scalfi_charge_2020}.

\subsubsection{Effect of the Thomas-Fermi screening length}
\label{sec:results:reciprocal:ltf}

\begin{figure}[!ht]
\center
\includegraphics[width=5.5in]{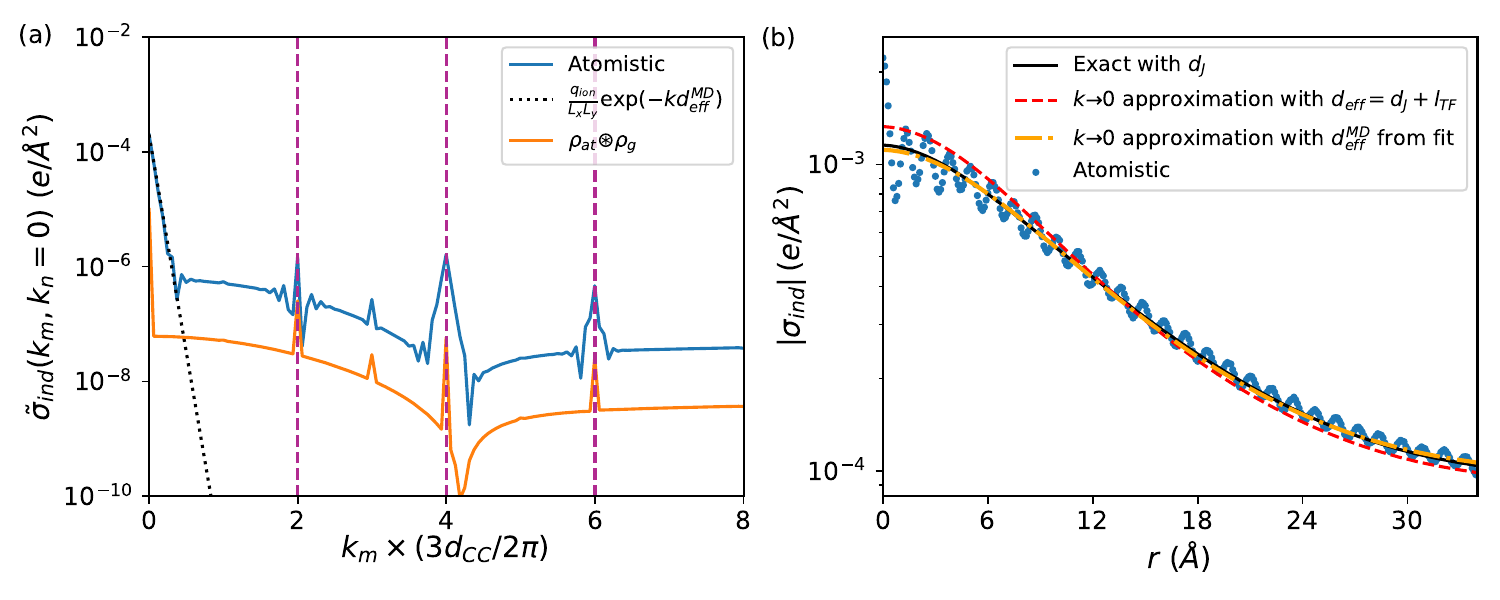}
\caption{
(a) Real part fo the Fourier coefficients of the two-dimensional induced charge density, defined in Eq.~\ref{eq:tildesigma}, for $z_{ion}=7.8$~\AA, $l_{TF}=5$~\AA\, and $k_n=0$.
As in panel~\ref{fig:fourier}a, the wavector is rescaled by $2\pi/b$, with $b$ the relevant dimension of the unit cell of the graphite lattice. The results from the atomistic description (blue solid line) are compared with the continuum approximation Eq.~\ref{eq:tildesigma:kzero} with $d$ fitted in the $k\to0$ region to extract an effective distance $d_{eff}^{MD}$ (dotted lines). We also show (orange line) the Fourier coefficients of the convolution of the atomic number density $\rho_{at}$ with the Gaussian charge density $\rho_g$ for each atom (see text). The vertical dashed lines indicate even values of the abscissa.
(b) Radial distribution of the induced charged density $\sigma_{ind}(r)$ (see Eq.~\ref{eq:rad}) for the same system. Results from the atomistic description (symbols) are compared with the continuum predictions using the exact 2D density Eq.~\ref{eq:ana} (black solid line), or the small $k$ approximation Eqs.~\ref{eq:anaapprox}-\ref{eq:deff}, using $d_{eff}=d_J+l_{TF}$ (red dashed line) or from the fit in panel (a) of the atomistic results (orange dashed-dotted line).
}
\label{fig:fourier:ltf}
\end{figure}

We now turn to the case of a finite Thomas-Fermi screening length. Fig.~\ref{fig:fourier:ltf}a shows the Fourier coefficients for $z_{ion}=7.8$~\AA\, and $l_{TF}=5$~\AA, for $k_n=0$. The results display the same features as in the $l_{TF}=0$ case. The peaks and baseline are again consistent with the convolution of the atomic density with the Gaussian distribution assigned to each atom. The decay for small $k$ is well described by an exponential, even though this behaviour is expected to be only approximate for finite $l_{TF}$ (see Eq.~\ref{eq:tildesigma:kzero}) because it assumes $k\ll \epsilon_1 k_{TF}/\epsilon_2=k_{TF}$ in the present case. The corresponding effective decay length, obtained by fitting the data by Eq.~\ref{eq:tildesigma:kzero} for $k\to0$, is $d_{eff}^{MD}=12.3\pm0.2$~\AA. Since the results of Section~\ref{sec:results:radial:ltf} suggest that the relevant ion-surface distance is that from the Jellium edge, we should compare this value with the prediction of Eq.~\ref{eq:deff} using $d=d_J$, which yields in that case $d_{eff}=11.1$~\AA. Even though the agreement is not perfect, this confirms the relevance of the continuum picture to describe the behaviour for small wavevectors, \textit{i.e.} large distances.

This can be further illustrated in real space with the radial distribution $\sigma_{ind}(r)$. In Fig.~\ref{fig:fourier:ltf}b, the results from the atomistic description are compared with the continuum predictions using the exact 2D density Eq.~\ref{eq:ana}, as well as the small $k$ approximation Eqs.~\ref{eq:anaapprox}-\ref{eq:deff}, using $d_{eff}=d_J+l_{TF}$ and from the fit in panel (a) of the atomistic results. As already discussed in Section~\ref{sec:results:radial:ltf}, the full solution with $d=d_J$ provides an excellent agreement with the atomistic results, except for the oscillations arising from the lattice structure. Note that compared \textit{e.g.} with Fig.~\ref{fig:radial:ltf}a, the $r$ range is significantly extended and the semi-logarithmic representation highlights the accuracy of the continuum prediction for large distances. The small $k\to0$ approximation with $d_{eff}=d_J+l_{TF}$ is also in relatively good agreement with the atomistic results, even though it overestimates the charge at short distance and underestimates it at large distance. Surprizingly, the small $k\to0$ approximation with $d_{eff}^{MD}$ fitted from the Fourier coefficients at small $k$ is in almost quantitative agreement with the exact result over the whole $r$ range and not only for large distances corresponding to small wavevectors.

\begin{figure}[!ht]
\center
\includegraphics[width=5.5in]{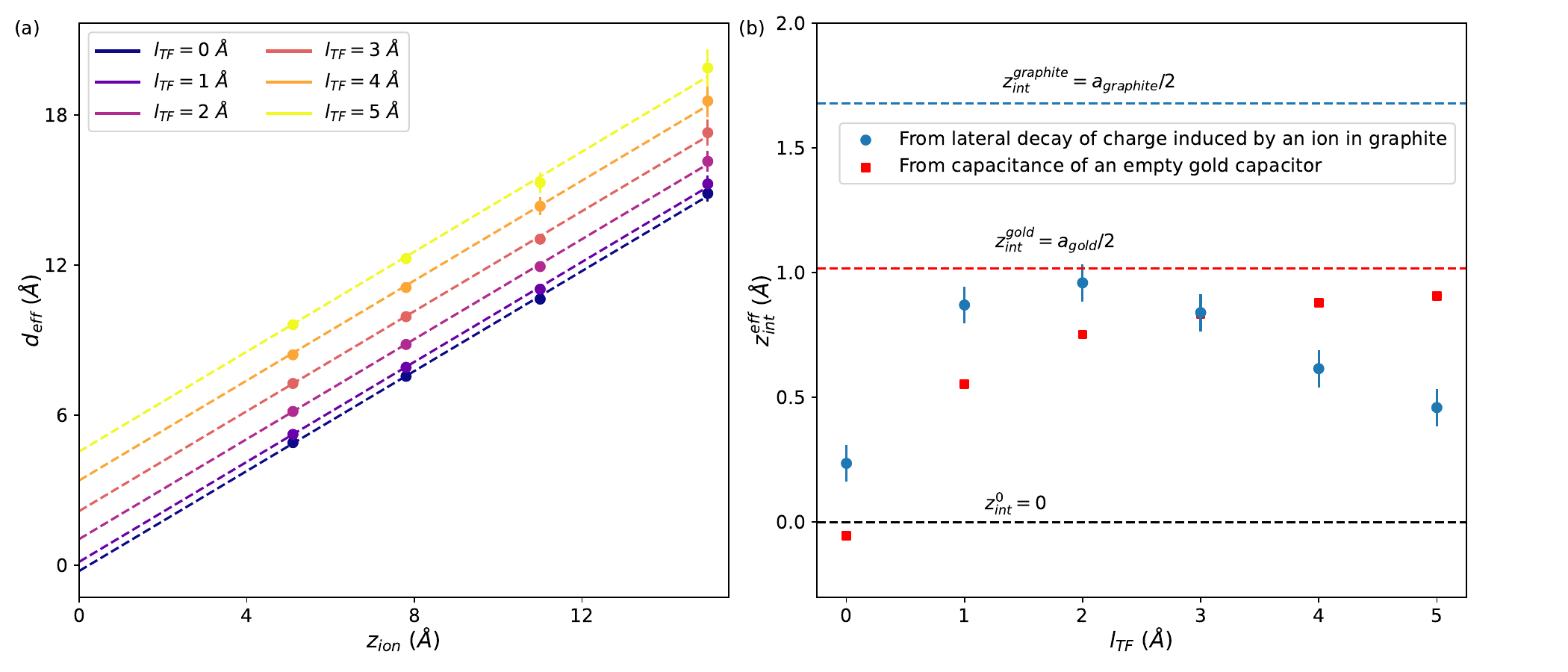}
\caption{
(a) Effective decay length $d_{eff}^{MD}$ obtained by fitting the $k\to0$ behaviour of the Fourier coefficients of the induced charge density (see Figs.~\ref{fig:fourier} and~\ref{fig:fourier:ltf}), as a function of the position of the ion $z_{ion}$, for $l_{TF}$ ranging from 0 to 5~\AA. The numerical results (symbols) are fitted to $d_{eff}=z_{ion}-z_{int}^{eff}+l_{TF}$ (dashed lines). (b) Effective position of the interface resulting from the fits of panel (a), as a function of the Thomas-Fermi screening length $l_{TF}$ as symbols in panel (blue circles). Also shown in panel (b) is the same quantity deduced from the effective screening length introduced in Ref.~\cite{Scalfi_tfmodel_2020} to account for the capacitance of empty capacitors with gold electrodes (red squares). The position of the Jellium edge is indicated as dashed horizontal lines for both graphite and gold electrodes.
}
\label{fig:fourier:leff}
\end{figure}

These observations support the idea that the finite $l_{TF}$ case can be well approximated by the solution for a perfect metal, using an effective distance (in particular in Eq.~\ref{eq:sigmaindltfzero} for a single charge) that depends on $l_{TF}$, reasonably described by Eq.~\ref{eq:deff}. In order to further test this approximation for $d_{eff}$, we report it in Fig.~\ref{fig:fourier:leff}a as a function of the ion position $z_{ion}$ for $l_{TF}$ between 0 and 5~\AA. The lines indicate fits of the form $d_{eff}=z_{ion}-z_{int}^{eff}+l_{TF}$, which provide in each case an effective position of the interface consistent with the linear scaling with both $z_{ion}$ and $l_{TF}$.

This quantity is then shown in Fig.~\ref{fig:fourier:leff}b as a function of $l_{TF}$. Consistently with the above results, $z_{int}^{eff}$ is found to be between $z_{int}^{0}$ and $z_{int}^{J}=a/2$, with a value close to $z_{int}^{0}$ for $l_{TF}=0$. In Ref.~\cite{Scalfi_tfmodel_2020} introducing the Thomas-Fermi model in constant-potential simulations, Scalfi \emph{et al.} considered the capacitance of empty gold capacitors with varying $l_{TF}$. They analyzed the results in terms of an equivalent circuit model consisting of capacitors in series, with capacitances per unit area $\epsilon_0/L_{vac}$ for the vacuum slab and $\epsilon_0/l_{eff}$ for each electrode, where the width $L_{vac}$ was taken as the distance between the two Jellium edges. The resulting effective screening length was found to coincide with $l_{TF}$ for large screening lengths, but to deviate when the latter become comparable to the atomic distances within the electrode. In Fig.~\ref{fig:fourier:leff}b, we also report the same results of Ref.~\cite{Scalfi_tfmodel_2020} by adopting a different perspective, namely assuming that the capacitance inside the metal can be estimated using $l_{TF}$ and that the capacitance of the whole system provides instead an effective vacuum slab width, which we translate in terms of the position of the interface. While the results are not identical with those obtained in the present work from the lateral decay of the charge induced by an ion, they share some similar features, such as being within half of the interplane distance from the first atomic plane, with a value close to $z_{int}^{0}$ for $l_{TF}=0$. We reiterate that the definition of $z_{int}^{eff}$ in the present work is based on an approximation only valid for $k\to0$. In addition, it has been suggested in Ref.~\cite{scalfi_charge_2020} that the effective distance between electrodes necessary to describe the capacitance might also be related to other parameters such as the width of the Gaussian charge distributions around each atom. Nevertheless, this comparison for different systems and different definitions of the location of the interface lead to overall consistent conclusions.

\section{Conclusion}

We have studied the charge induced in a Thomas-Fermi metal by an ion in vacuum, using an atomistic description employed in constant-potential molecular dynamics simulations, and compared the results with the (semi-)analytical predictions obtained from continuum electrostatics. Following a previous study for a gold electrode described as a perfect metal (with a Thomas-Fermi screening length $l_{TF}=0$), with an ion in vacuum and in water, we considered the effect of screening inside the metal by restricting ourselves to the case of an ion in vacuum near a graphite electrode, varying $l_{TF}$ and the position of the ion with respect to the surface. Our analysis of the induced charge included not only 2D density maps and their radial averages, but also their Fourier coefficients in order to extract the long-range behaviour under the periodic boundary conditions inherent to molecular simulations. The comparison with the continuum picture also included new analytical results taking these periodic boundary conditions into account.

Overall, the continuum predictions capture most of the features observed with the atomistic description (except the oscillations due to the discrete atomic sites of the graphite lattice, as expected), provided that the distance of the ion from the surface and the Thomas-Fermi length are larger than the inter-atomic distances within the electrode. The analysis of the Fourier coefficients for small wavevectors shows that, at large distance, the finite $l_{TF}$ case can be well approximated by the solution for a perfect metal, using an effective distance, reasonably described by the sum of the distance between the ion and the Jellium edge (which corresponds to the electrode/vacuum interface) and the Thomas-Fermi screening length. Surprisingly, we find that this approximation is in fact rather accurate even at intermediate and short distances. The linear relation between the effective distance and both the ion position and the screening length can also be translated in terms of an effective position of the interface, which depends on $l_{TF}$, is within half of the interplane distance from the first atomic plane of the electrode, and is broadly consistent with another definition from the capacitance of empty capacitors. 

These conclusions support in particular an early analytical prediction by Vorotyntsev and Kornyshev for a single charge (nonperiodic case) located at the interface between a Thomas-Fermi metal and a polarizable medium. We are currently exploring the effect of the interplay between screening within the metal and by a polar solvent on the charge induced by an ion, using molecular simulations. Further studies could also include the interplay with screening by ions in an electrolyte, in order to provide simple analytical approximations in that case, which is the most relevant for practical applications of electrode/electrolyte interfaces.

\section*{Acknowledgements}

We dedicate this article to Giovanni Ciccotti on the occasion of his 80$^{th}$ birthday. In particular, BR is grateful for many years of friendship and scientific or other discussions, spiced with his strong opinions expressed in Italian, English or French. We would like to thank Laura Scalfi, Mathieu Salanne, as well as David Limmer (with the financial support of the France-Berkeley Fund) for useful discussions.
This project was supported by the European Research Council under the European Union’s Horizon 2020 research and innovation program (project SENSES, grant Agreement No. 863473). The authors acknowledge access to HPC resources from GENCI-IDRIS (grant no. 2022-AD010912966R1).


\section*{Appendix}
\label{sec:appendix}

\begin{figure}[!ht]
\center
\includegraphics[width=5in]{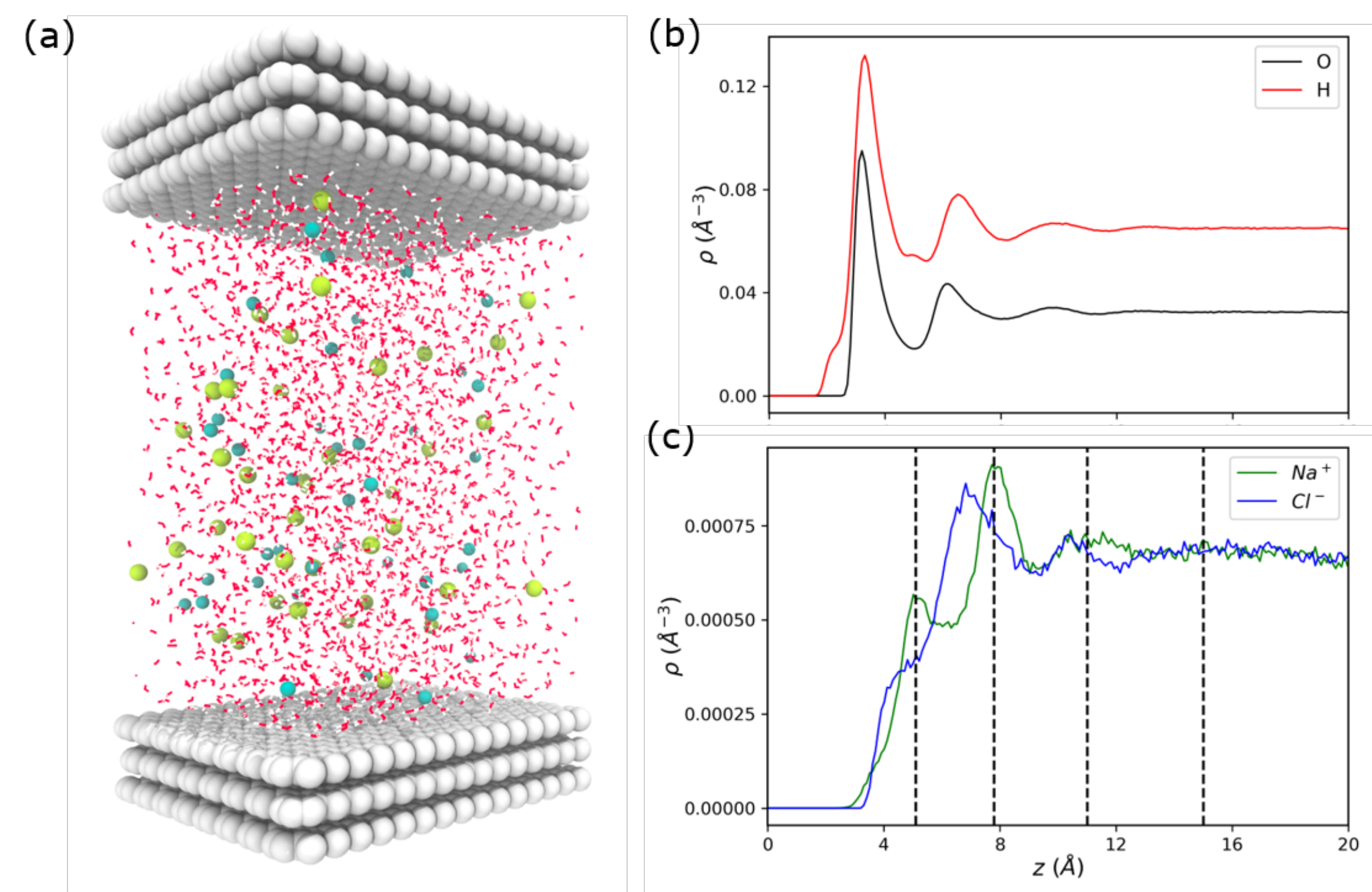}
\caption{
(a) Simulated system for a 1~M NaCl aqueous solution between graphite electrodes.
Equilibrium density profiles for water and ions are shown in panels (b) and (c), respectively. The vertical dashed lines in panel (c) indicate the selected positions for the present study of an ion in vacuum.
}
\label{fig:appendix}
\end{figure}

As indicated in Section~\ref{sec:systemmethod}, the ionic positions $z_{ion}$ considered in the present work were chosen based on the equilibrium density profile of Na$^{+}$ at the interface between graphite and a 1~M NaCl aqueous solution using the same model. The corresponding simulations were performed with the force field described in the main text and the SPC/E water model~\cite{berendsen_missing_1987}, with the system illustrated in Fig.~\ref{fig:appendix}a. It contains 2 graphite electrodes (with $l_{TF}=0)$, each consisting of 1440 atoms in 3 planes, maintained at the same potential (\emph{i.e.} a voltage $\Delta\psi=0$~V), as well as 2160 water molecules and 39 ion pairs. The box dimensions are $L_x = 34.10$~\AA, $L_y = 36.91$~\AA\, and $L_z = 211.67$~\AA, and periodic boundary conditions are applied only in the $x$ and $y$ directions. Newton's equations of motion are integrated using a time step of 1~fs. The system is first equilibrated during 0.5~ns in the $NPT$ ensemble at a pressure of 1 atm by applying a corresponding force on the electrodes (pistons) and a temperature $T=298$~K by applying a Nos\'e-Hoover chain thermostat with a time constant of 1~ps. The distance between the first atomic plane of the electrodes is then fixed (to 56.268~\AA) for subsequent runs in the $NVT$ ensemble. After 0.5~ns of equilibration, a production run of 8~ns is performed, sampling the configurations every 1~ps to compute the equilibrium density profiles of water oxygen and hydrogen atoms (see Fig.~\ref{fig:appendix}b) and sodium and chloride ions (see Fig.~\ref{fig:appendix}c). The vertical dashed lines in panel Fig.~\ref{fig:appendix}c indicate the selected positions for the present study of an ion in vacuum.

\vspace{1cm}







\newpage
\bibliographystyle{unsrt} 

\end{document}